\documentclass[aps,prd,twocolumn,groupedaddress]{revtex4-1}
\usepackage{graphicx}
\usepackage{wrapfig}
\usepackage{latexsym}
\usepackage{amssymb}
\usepackage{amsfonts,color}
\usepackage{float}
\usepackage{amsmath}
\begin{document}


\title{An Electrostatic Analogy for Symmetron Gravity}
\author{Lillie Ogden$^{(1)}$, Katherine Brown$^{(1)}$, Harsh Mathur$^{(2)}$} 
\author{Kevin Rovelli$^{(1)}$}
\affiliation{$^{(1)}$Department of Physics, Hamilton College, Clinton, NY 13323}

\vspace{1mm}
\affiliation{$^{(2)}$Department of Physics, Case Western Reserve University, Cleveland, Ohio 44106-7079}


\date{\today}

\begin{abstract}
The symmetron model is a scalar-tensor theory of gravity with a screening 
mechanism that suppresses the effect of the symmetron field at high densities 
characteristic of the solar system and laboratory scales 
but allows it to act with gravitational strength at low density on the cosmological scale. 
We elucidate the screening mechanism by showing that in the quasi-static Newtonian
limit there are precise analogies between symmetron gravity and electrostatics
for both strong and weak screening. For strong screening 
we find that large dense bodies behave in a manner analogous to perfect conductors
in electrostatics. Based on this analogy we find that the symmetron field exhibits a lightning rod effect
wherein the field gradients are enhanced near the ends of pointed or elongated objects.
An ellipsoid placed in a uniform symmetron gradient is shown to experience a torque.
By symmetry there is no gravitational torque in this case. Hence this effect 
unmasks the symmetron and might
serve as the basis for future laboratory experiments. 
The symmetron force between a point mass and a large dense body includes
a component corresponding to the interaction of the point mass with its image in the larger body.
None of these effects have counterparts
in the Newtonian limit of
Einstein gravity. We discuss the similarities between symmetron gravity and the chameleon model
as well as the differences between the two. 

\end{abstract}


\maketitle

\section{Introduction \label{introduction}}

In 1961 Brans and Dicke introduced a scalar-tensor theory of gravity as a consistent
alternative to general relativity \cite{brans, will}. More recently 
the puzzle of dark energy and the accelerated expansion of the
Universe has provided a fresh impetus to study alternatives to general relativity \cite{supernova}. 
Moreover models of fundamental physics with extra dimensions generally lead to scalar-tensor gravity 
rather than general relativity as the effective four dimensional theory of gravity
\cite{trodden}. Thus the study of scalar-tensor gravity remains a compelling subject. 
In a scalar-tensor theory the Einstein-Hilbert action is 
unchanged from general relativity but matter fields are assumed to be coupled to a Jordan frame
metric $\tilde{g}_{\mu \nu}$ which is related to the Einstein metric $g_{\mu \nu}$ by
$\tilde{g}_{\mu \nu} = A^2(\phi) g_{\mu \nu}$ where the conformal factor $A$ depends on the
scalar field $\phi$. Different scalar-tensor theories are distinguished by the form of the conformal
factor $A(\phi)$ and the choice of the potential $V(\phi)$ for the scalar Lagrangian. 

Scalar tensor theories in which the scalar field is governed by a symmetry breaking potential
were studied in 1992 by Gessner \cite{gessner} and Dehnen {\em et al.} \cite{dehnen} as a possible explanation
of the flatness of rotation curves. These works are an important precursor to the symmetron model proposed
in 2006 by Hinterbichler and Khoury \cite{hk} as a model of dark energy. 
For a modified theory of gravity to have significant effects
on the cosmological scale and yet remain consistent
with solar system and laboratory tests of general relativity it is necessary to 
devise a suitable screening mechanism. The chameleon model
introduced by Khoury and Weltman \cite{kw} 
was the first scalar-tensor theory with a screening mechanism
that suppressed the effects of the 
scalar in high density
environments like the Earth while allowing it to play the role of
dark energy on the cosmological scale. Subsequently
Hinterbichler and Khoury introduced the symmetron model \cite{hk} 
wherein the scalar field undergoes a density-driven symmetry breaking
phase transition and is thereby screened in high density environments.
Jones-Smith and Ferrer \cite{jsf} elucidated the screening mechanism
of the chameleon model by showing that the screening of the chameleon
field by large dense bodies was analogous to the screening of electric
fields by perfect conductors. In ref \cite{jsf} it was mentioned but not shown 
that a similar analogy exists between the symmetron field and electrostatics. 
The purpose of this paper is to make the analogy explicit. 
Our results presumably also apply to related variants of the symmetron 
model, for example \cite{variants}. 


The principal findings of ref \cite{jsf} were that in the quasi-static Newtonian limit
the equations of chameleon gravity were analogous to different electrostatic problems
for both strong and weak screening. In particular large dense bodies were shown to
be essentially equipotentials of the chameleon field and to behave in a manner
precisely analogous to perfect conductors in electrostatics. Although there are
important differences between chameleon and symmetron gravity, and the detailed
arguments are different in the two cases, we find nonetheless that in the
quasi-static Newtonian limit the equations of symmetron gravity also map
onto different electrostatic problems in the limits of strong and weak screening. 
Hence symmetron gravity also shows a range of remarkable phenomena
that are present in the chameleon
but completely absent in Newtonian gravity. Notably symmetron
gravity exhibits a lightning rod effect wherein the field gradients become
concentrated near the ends of a pointed or elongated object much as 
electric and chameleon fields do \cite{jsf}. Also by analogy to electrostatics 
and the chameleon, an ellipsoid placed in 
uniform symmetron gradient experiences a torque tending to align
the major axis with the field gradient. This effect is remarkable
because generally Newtonian gravity overwhelms screened symmetron effects
but in this circumstance the gravitational torque cancels perfectly
by symmetry thereby unmasking the symmetron. Finally a small dense body brought
near a much larger dense body will be repelled by its image mass much as
point charges are attracted to their images in a perfect conductor. 
Whether the net force can become repulsive is an intriguing 
question discussed further below.
The chameleon counterpart of this effect was first pointed out in ref \cite{afshordi}. 
Remarkably the shape dependence of chameleon and symmetron gravity noted above
have also been found to be exhibited by other modified gravity models based on the Vainshtein mechanism \cite{davis}. 



The chameleon and symmetron models are amenable to experimental tests on scales ranging from the
laboratory to cosmology. Constraints on the symmetron from cosmology have been studied in 
ref \cite{matas} and on the laboratory scale by ref \cite{upadhye}. Laboratory experiments have
traditionally relied on torsion oscillators \cite{adelberger} but recently atom interferometry has been used to 
place constraints on the chameleon \cite{mueller,hinds,burrage} and in principle can be used to place
constraints on the symmetron as well. For a review of constraints from laboratory and astrophysical  
observations from the symmetron, chameleon, and related models see ref \cite{trodden, burragereview}.
Very recently the symmetron model has been reapplied \cite{millington} to the dark matter
problem of galaxy rotation, the context in which it was first introduced \cite{gessner}, 
stimulated by the measurement of a correlation between the observed and baryonic acceleration in
disk galaxies \cite{stacy}.
It is hoped that the qualitative understanding provided by the electrostatic analogy and the effects
suggested by it will help sharpen experimental tests e.g. by suggesting geometries in which the 
symmetron effects are enhanced or unmasked as in the case of the ellipsoid torque noted above.

The paper is organized as follows. In section II we introduce
the symmetron model and analyze and review some simple geometries.
These include a dense sphere and its interaction with a test mass studied
by \cite{hk}; a semi-infinite slab which provides insights into surface forces 
and the field profile in the boundary layer in the simplest possible circumstance;
and a finite slab which provides a simple soluble example of domain walls that can
arise in the symmetron model \cite{levon, mota1, mota2}. The first two geometries motivate
the formulation of the electrostatic analogy; the third geometry exemplifies an effect that is present 
in the symmetron model but has no counterpart for the chameleon. We present
the symmetron electrostatic analogy in section \ref{sec:analogy}.
Applications of the analogy to some of the same geometries that were discussed in connection 
with the chameleon in ref \cite{jsf} are relegated to Appendix \ref{sec:applications}. 
Among other results we present the force
on a dense sphere placed in a uniform symmetron field gradient in both the screened and
unscreened limits; the field of an ellipsoid demonstrating the lightning rod effect in context of
both the near and far field behavior; and the torque on an ellipsoid placed in a uniform
symmetron gradient. We highlight the parallels and differences between the symmetron and chameleon
for these geometries and in these cross comparisons 
we provide some results that were omitted for reasons of
space in ref \cite{jsf}. In section \ref{sec:image} we analyze the interaction between
a small dense body and a large dense sphere using the electrostatic analogy and the 
method of images. 
This analysis suggests the intriguing possibility that under appropriate
conditions the net symmetron force between the two objects can become negative. Further
work along the lines discussed in section \ref{sec:image} is needed in order to 
conclusively demonstrate that the net symmetron force can indeed become negative.


\section{The Symmetron Model and the Thin Shell effect}

\label{sec:simple}


\subsection{The Model}

In the quasi-static Newtonian limit of general relativity \cite{weinberg}, 
the gravitational field may be described by the Newtonian
scalar potential $\psi$ which obeys Poisson's equation
\begin{equation}
\nabla^2 \psi = \frac{1}{2 M_{{\rm P}}^2} \rho.
\label{eq:newtonianpot}
\end{equation}
Here $M_{{\rm P}} = 1/\sqrt{8 \pi G}$ and $\rho$ denotes the density of matter that sources the
gravitational field. We work in a system of units where $\hbar = c = 1$. 
In a scalar-tensor theory of gravity, such as the symmetron model, gravity has an additional scalar degree
of freedom. In the quasi-static Newtonian limit 
the additional scalar field $\phi$ satisfies
\begin{equation}
\nabla^2 \phi = \frac{d}{d \phi} V_{{\rm eff}} (\phi).
\label{eq:scalar}
\end{equation}
The effective potential $V_{{\rm eff}} = V(\phi) + \rho A(\phi)$. 
In the symmetron model \cite{hk} the potential $V$ is taken to have the standard symmetry breaking form
\begin{equation}
V(\phi) = - \frac{1}{2} \mu^2 \phi^2 + \frac{1}{4} \lambda \phi^4
\label{eq:symmetrybreakingpot}
\end{equation}
and the conformal factor is taken to be
\begin{equation}
A(\phi) = 1 + \frac{1}{2 M^2} \phi^2.
\label{eq:conformal}
\end{equation}
$M, \mu$ and $\lambda$ are parameters of the symmetron model. 
Given the mass distribution $\rho$ in principle one can determine the symmetron field
by solving eq (\ref{eq:scalar}). 

The symmetron field manifests itself by exerting forces on matter. 
A test particle of mass $m_0$ moving non-relativistically obeys
\begin{equation}
m_0 \frac{d {\mathbf v}}{d t} =  - m_0 \nabla \psi - m_0 \left( \frac{\partial A}{\partial \phi} \right) \nabla \phi.
\label{eq:testparticle}
\end{equation}
The first term on the right hand side of eq (\ref{eq:testparticle}) is the Newtonian gravitational force;
the second term is the additional ``fifth-force'' due to the symmetron field. 

In order to gain some intuition into the symmetron model it is helpful to
solve the field eq (\ref{eq:scalar}) under various simple circumstances. First suppose $\rho = 0$. 
In this case $V_{{\rm eff}} = V$ and the effective potential is
a double well with minima at $\pm \phi_0$ where $\phi_0 = \mu/\sqrt{\lambda}$. The symmetron will then
spontaneously break symmetry. At low energy let us assume that $\phi \approx \phi_0 + \xi$. Expanding
$V$ about the minimum to second order we find that eq (\ref{eq:scalar}) simplifies to
\begin{equation}
\nabla^2 \xi = 2 \mu^2 \xi.
\label{eq:vacuum}
\end{equation}
Thus within our approximation we find that in empty space the deviation of the symmetron field
from its vacuum value $\phi_0$ is governed by massive electrostatics with mass scale
$\sqrt{2} \mu$. 

Another simple circumstance is to suppose that space is uniformly filled with a 
homogeneous fluid with density $\rho$. So long as $\rho/M^2 > \mu^2$, then $V_{{\rm eff}}$ has a single
minimum at $\phi = 0$. Assuming that the field $\phi$ remains close to the minimum we may linearize the
equations of motion to obtain
\begin{equation}
\nabla^2 \phi = \left( \frac{\rho}{M^2} - \mu^2 \right) \phi.
\label{eq:dense}
\end{equation}
Thus in the dense phase too the symmetron field is governed by massive electrostatics. 
Assuming the density is sufficiently high, $\rho/M^2 \gg \mu^2$, 
the mass scale is approximately given by
$\sqrt{\rho}/M$. 


\subsection{Solution for a Dense Sphere}
\label{sec:exactsphere}

Following Hinterbichler and Khoury \cite{hk} 
we now briefly consider the symmetron field produced by a solid sphere of
radius $R$ and a uniform density $\rho$ immersed in a vacuum. 
The sphere might represent a planet, star, dwarf spheroidal galaxy
or a brass sphere in a vacuum chamber in a table top experiment. Physically we expect that far
from the sphere the symmetron field $\phi \rightarrow \phi_0$. The form of the field inside the sphere
depends on whether the sphere is large or small compared to the length scale $M/\sqrt{\rho}$.
Thus $\alpha = \rho R^2/M^2$ is a key parameter that characterizes the dense sphere.
In the thin shell regime ($\alpha \gg 1$)
we expect that deep in the interior of the sphere the symmetron will attain the
equilibrium value $\phi \rightarrow 0$ except 
in a thin shell of depth $M/\sqrt{\rho}$ near the surface of the sphere. 
In the opposite thick shell regime ($\alpha \ll 1$) $\phi$ is not able to attain the equilibrium value in the interior.

To calculate the profile we wish to solve eq (\ref{eq:vacuum}) in the exterior and
eq (\ref{eq:dense}) in the interior subject to the boundary conditions that
$\phi$ must be regular at the origin and $\phi \rightarrow \phi_0$ as $r \rightarrow \infty$. In addition
we require that $\phi$ and its radial derivative must be continuous across the surface of the sphere.
Imposing these conditions yields the solution
\begin{eqnarray}
\phi & = & \phi_0 + \frac{A}{r} \exp ( - \sqrt{2} \mu r ) \hspace{3mm} {\rm for} \hspace{3mm} r > R
\nonumber \\
& = & \frac{C}{r} \sinh \left( \sqrt{\alpha} \frac{r}{R} \right) \hspace{3mm} {\rm for} \hspace{3mm} r < R
\nonumber \\
\label{eq:exactprofile}
\end{eqnarray}
with coefficients
\begin{eqnarray}
A & = & \phi_0 R \frac{ (\sinh \sqrt{\alpha} - \sqrt{\alpha} \cosh \sqrt{\alpha} ) }{ (\sqrt{ \alpha } \cosh \sqrt{\alpha} 
+ \sqrt{2} \mu R \sinh \sqrt{\alpha} ) } \nonumber \\
C & = & \phi_0 R \frac{ (1 + \sqrt{2} \mu R ) }{ ( \sqrt{\alpha} \cosh \sqrt{\alpha} + \sqrt{2} \mu R \sinh \sqrt{\alpha} ) }.
\label{eq:ac}
\end{eqnarray}

In the thin shell approximation ($\alpha \gg 1$) eq (\ref{eq:exactprofile}) simplifies to
\begin{equation}
\phi \approx \phi_0 \frac{1}{\sqrt{\alpha} } \frac{R}{r} \exp \left[ - \frac{\sqrt{\alpha}}{R} (R - r) \right]
\label{eq:thinshellsphere}
\end{equation}
for $r < R$. 
Hence as anticipated $\phi \approx 0$ except for a thin shell near the surface. In the thick shell limit
($\alpha \ll 1$) 
\begin{equation}
\phi \approx \phi_0 \left( 1 - \frac{1}{2} \alpha + \frac{1}{6} \alpha \frac{r^2}{R^2} + \ldots \right)
\label{eq:thickshellsphere}
\end{equation}
for $r < R$. Hence as anticipated $\phi$ does not deviate much from its vacuum value $\phi_0$ in the interior of
the dense sphere in the thick shell limit. Intuitively, the dense sphere is much too small to cause a significant 
distortion of the symmetron field. 

In the thin shell limit ($\alpha \gg 1$) eq (\ref{eq:exactprofile}) simplifies to
\begin{equation}
\phi \approx \phi_0 - \phi_0 \left( 1 - \frac{1}{\sqrt{\alpha}} \right) \left( \frac{R}{r} \right) \exp \left[ - \sqrt{2} \mu (r - R) \right]
\label{eq:thinshelloutside}
\end{equation}
for $r > R$. 
For $\mu r \ll 1$ the exterior profile has the even simpler form
\begin{equation}
\phi \approx \phi_0 - \phi_0 \frac{R}{r}
\label{eq:thinshellsphereelectrostatic}
\end{equation}
a form reminiscent of the electrostatic potential of a conducting sphere. 
In the thick shell limit $(\alpha \ll 1)$ eq (\ref{eq:exactprofile}) simplifies to 
\begin{equation}
\phi \approx \phi_0 - \phi_0 \frac{\alpha}{3} \left( \frac{R}{r} \right) \exp \left[ - \sqrt{2} \mu ( r - R ) \right] 
\label{eq:thicksphereoutside}
\end{equation}
for $r > R$. Thus to a good approximation the field has its vacuum value $\phi_0$. At the surface of the
sphere the field is lower than the vacuum value by a factor of $1 - (\alpha/3)$. The small deviation from
the vacuum value decays exponentially away from the surface
on a length scale $1/\sqrt{2} \mu$. Again if we suppose $\mu r \ll 1$
the exterior profile has the even simpler form
\begin{equation}
\phi \approx \phi_0 - \phi_0 \frac{\alpha}{3} \frac{R}{r}.
\label{eq:thicksphereelectrostatic}
\end{equation}

Eqs (\ref{eq:testparticle}) and (\ref{eq:thicksphereelectrostatic})
reveal that in the thick shell limit a test particle of mass $m_0$ will experience a symmetron fifth force of magnitude
$F_\phi = m_0 \phi_0^2 \alpha R/(3 M^2 r^2)$ directed towards the center of the dense sphere. 
It is instructive to rewrite this expression in the form $F_\phi = (\phi_0^2/4 \pi M^4) m_0 m_S / r^2$ 
where $m_S$ is the mass of the dense sphere. In this form the similarity of the fifth force to Newtonian
gravity is apparent. In the thick shell regime the 
ratio of the symmetron force to the gravitational force is $ F_\phi/F_{{\rm grav}} = 2 ( \phi_0 M_P / M^2 )^2$. 
Following \cite{hk} we take the parameters of the
symmetron model to satisfy $\phi_0 M_P/M^2 \sim 1$; hence in the thick shell regime the fifth force is comparable in 
strength to gravity. In the thin shell regime by contrast it follows from eqs (\ref{eq:testparticle}) and 
(\ref{eq:thinshellsphereelectrostatic}) that the 
the magnitude of the symmetron force is $F_\phi = m_0 \phi_0^2 R/ M^2 r^2$ and the ratio of the symmetron
force to the gravitational force is
\begin{equation}
\frac{F_\phi}{ F_{{\rm grav}} } = 6 \left( \frac{ \phi_0 M_P }{ M^2 } \right)^2 \frac{1}{\alpha}.
\label{eq:thinshelleffect}
\end{equation}
Hence in the thin shell regime ($\alpha \gg 1$) 
the symmetron force is suppressed relative to gravity by a factor $1/\alpha$. 
Another remarkable characteristic of the symmetron force on the test mass in the thin shell regime
is that it is determined by the volume of the dense sphere rather than its mass in sharp contrast to the thick
shell regime or Newtonian gravity. 
For the chameleon too there is a thin shell suppression of the fifth force
on a test mass but there is no counterpart to the remarkable mass independence of the force obtained for the symmetron.


\subsection{Semi Infinite Slab}
\label{sec:semislab}

Another simple circumstance that is instructive to analyze is a semi-infinite slab. Suppose
that the left half-space $x < 0$ is filled with a fluid of density $\rho$ and the right half-space,
$x > 0$, is empty. Intuitively we expect that $\phi \rightarrow \phi_0$ as $x \rightarrow \infty$
and $\phi \rightarrow 0$ as $x \rightarrow - \infty$. The ratio of the mass scale in the dense body
to the mass scale in the vacuum is the parameter $\beta = \sqrt{\rho}/{M \mu}$. We confine
our attention to the thin shell regime $\beta \gg 1$. 

As we will see below, in one dimension it is necessary to solve the exact
field equation in the exterior of the slab. Thus for $x > 0$ we must solve
\begin{equation}
\frac{d^2 \phi}{d x^2} = \frac{d}{d \phi} V = \frac{d}{d \phi} \left[ \frac{\lambda}{4} \phi^4 - \frac{\mu^2}{2} \phi^2 \right]
\label{eq:1dvaceq}
\end{equation}
subject to the boundary condition that $\phi \rightarrow \phi_0$ as $x \rightarrow \infty$.
Eq (\ref{eq:1dvaceq}) 
has the simple exact solution \cite{tanmay}
\begin{equation}
\phi = \phi_0 \tanh \left[ \frac{ \mu }{\sqrt{2}} ( x + x_0 ) \right]
\label{eq:kink}
\end{equation}
where $x_0$ is an arbitrary constant. In the interior of the dense body it is sufficient to solve
eq (\ref{eq:dense}) which has the solution
\begin{equation}
\phi = A \exp \left( \frac{\sqrt{\rho}}{M} x \right)
\label{eq:slabin}
\end{equation}
for $x < 0$. Here we have made use of the boundary condition $\phi \rightarrow 0$ 
as $x \rightarrow - \infty$. Finally we must impose the matching conditions that $\phi$ and
$d\phi/d x$ should be continuous at the boundary of the slab. Imposing the matching condition
and assuming $\beta \gg 1$ we obtain
\begin{eqnarray}
\phi & = & \phi_0 \frac{1}{\sqrt{2}} \frac{1}{\beta} \exp \left( \frac{\sqrt{\rho}}{M} x \right) \hspace{2mm}
{\rm for} \hspace{2mm} x < 0 
\nonumber \\
& = & \phi_0 \tanh \left[ \frac{1}{\sqrt{2}}( \mu x + \frac{1}{\beta} ) \right] \hspace{2mm} {\rm for} \hspace{2mm}
x > 0.
\label{eq:slabsol}
\end{eqnarray}
From eq (\ref{eq:slabsol}) we see that $\phi$ is of order $\phi_0/\beta$ near the surface of the slab and it
rises to a value close to $\phi_0$ only for $x \gg 1/\mu$. Because the field approaches the vacuum value only
on the long length scale $1/\mu$ we see {\em a posteriori} that it was necessary to solve the field equations
exactly in the exterior. 

The dense body experiences a pressure due to the symmetron 
field which we can calculate as follows. Making use of eq (\ref{eq:testparticle})
it follows that the $x$-component of the force per unit area on the slab is given by
\begin{equation}
{\cal S}_x = - \frac{\rho}{M^2} \int_{-\infty}^0 d x \; \phi \frac{d}{dx} \phi.
\label{eq:pressure}
\end{equation}
Assuming that the field in the interior of the slab has the exponential form
in eq (\ref{eq:slabin}) this works out to
\begin{equation}
{\cal S}_x = - \frac{1}{2} \left( \frac{d \phi}{d x} \right)^2 \mid_{x \rightarrow 0}. 
\label{eq:thinpressure}
\end{equation}
This result will be helpful below in developing a more general expression for the force on a dense
body in the thin shell limit. Making use of eq (\ref{eq:slabsol}) and eq (\ref{eq:thinpressure}) we obtain the
explicit result
\begin{equation}
{\cal S}_x = - \frac{1}{2} ( \mu \phi_0 )^2. 
\label{eq:thinpressureexplicit}
\end{equation}
Remarkably the pressure has a universal form, independent of the density of the slab. We will see
below that this is a generic feature of the thin shell limit of the symmetron model.


\subsection{Domain Walls}

Models that break a discrete Z$_2$ symmetry support topological defects called domain walls \cite{tanmay}.
The properties of domain walls in symmetron models were studied in \cite{levon}.
We begin our analysis by considering
slabs of finite thickness in order to gain some insight into the domain
walls formed by the symmetron field. 
Suppose that the region
$| x | < L$ is filled with a fluid of density $\rho$. The space on either side of the slab ($x > L$ or 
$x < -L$) is empty. Since the symmetron has two identical vacua $\phi = \pm \phi_0$ it follows
that there are two circumstances to consider. The first possibility the same vacuum is to be found on either
side of the slab. We may take $\phi \rightarrow \phi_0$ for $x \rightarrow \pm \infty$. This circumstance
leads to a symmetric profile for the field $\phi$. The other possibility is that different vacua are found
on the two sides of the slab. To be definite we may take $\phi \rightarrow \pm \phi_0$ as 
$x \rightarrow \pm \infty$. In this case the symmetron profile is anti-symmetric and we may regard
the slab as a domain wall. 

In the symmetric case we wish to solve eq (\ref{eq:1dvaceq}) in the
exterior of the slab and eq (\ref{eq:dense}) in the interior. We impose the boundary condition $\phi \rightarrow
\phi_0$ for $x \rightarrow \pm \infty$. Since the boundary condition is symmetric we expect a symmetric solution
\begin{eqnarray}
\phi & = & C \cosh \left( \frac{ \sqrt{\rho} x }{M } \right) \hspace{2mm} {\rm for} \hspace{2mm} 0 < x < L,
\nonumber \\
& = & \phi_0 \tanh \left[ \frac{ \mu }{\sqrt{2}} (x + x_0) \right] \hspace{2mm} {\rm for} \hspace{2mm} x > L.
\label{eq:symmetricsol}
\end{eqnarray}
Here $C$ and $x_0$ are arbitrary constants. and the solution is assumed to be a symmetric function of $x$.
By matching $\phi$ and its derivative at $x = L$ we can determine the arbitrary constants $C$ and $x_0$ and
obtain the explicit solution
\begin{eqnarray}
\phi & = & \phi_0 \frac{ \tanh \Delta_S }{ \cosh ( \sqrt{\rho} L/M ) } \cosh ( \sqrt{\rho} x/ M) 
\hspace{2mm} {\rm for} \hspace{2mm} 0 < x < L, 
\nonumber \\
& = & \phi_0 \tanh \left[ \frac{ \mu }{\sqrt{2} } (x - L) + \Delta_S \right] \hspace{2mm} {\rm for} \hspace{2mm} x > L.
\label{eq:symmetricexact}
\end{eqnarray}
The phase shift $\Delta_S$ is given by $\tanh \Delta_S = \sqrt{1 + f_S^2} - f_S$ with
\begin{equation}
f_S = \frac{ \sqrt{\rho}}{\mu M \sqrt{2}} \tanh \left( \frac{ \sqrt{\rho} L}{M} \right).
\label{eq:fs}
\end{equation}

The antisymmetric case may be analyzed similarly. We wish
to solve eqs (\ref{eq:1dvaceq}) in the exterior and eq (\ref{eq:dense}) in the interior. We impose
the boundary conditions $\phi \rightarrow \pm \phi_0$ for $x \rightarrow \pm \infty$. In view of the antisymmetry
of the boundary conditions we make the ansatz
\begin{eqnarray}
\phi & = & C \sinh \left( \frac{ \sqrt{ \rho} x }{ M } \right) \hspace{2mm} {\rm for} \hspace{2mm} 0 < x < L,
\nonumber \\
& = & \phi_0 \tanh \left[ \frac{ \mu }{\sqrt{2}} (x + x_0) \right] \hspace{2mm} {\rm for} \hspace{2mm} x > L.
\label{eq:antisymmetricsol}
\end{eqnarray}
Here $C$ and $x_0$ are arbitrary constants and the solution is assumed to be an antisymmetric 
function fo $x$. By imposing the continuity of $\phi$ and its derivative at $x = L$ we can determine the
arbitrary constants and obtain the explicit solution
\begin{eqnarray}
\phi & = & \phi_0 \frac{ \tanh \Delta_A }{\sinh ( \sqrt{ \rho } L/M ) } \sinh ( \sqrt{ \rho } x / M )
\nonumber \\
& = & \phi_0 \tanh \left[ \frac{\mu}{\sqrt{2}} ( x - L ) + \Delta_A \right].
\label{eq:antisymmetricexact}
\end{eqnarray}
The phase shift is given by $\tanh \Delta_A = \sqrt{ 1 + f_A^2} - f_A$ with
\begin{equation}
f_A = \frac{ \sqrt{\rho}}{\mu M \sqrt{2}} \coth \left( \frac{ \sqrt{\rho} L}{M} \right).
\label{eq:fa}
\end{equation}

\begin{figure}[!ht]
\begin{center}
\includegraphics[width=0.45\textwidth]{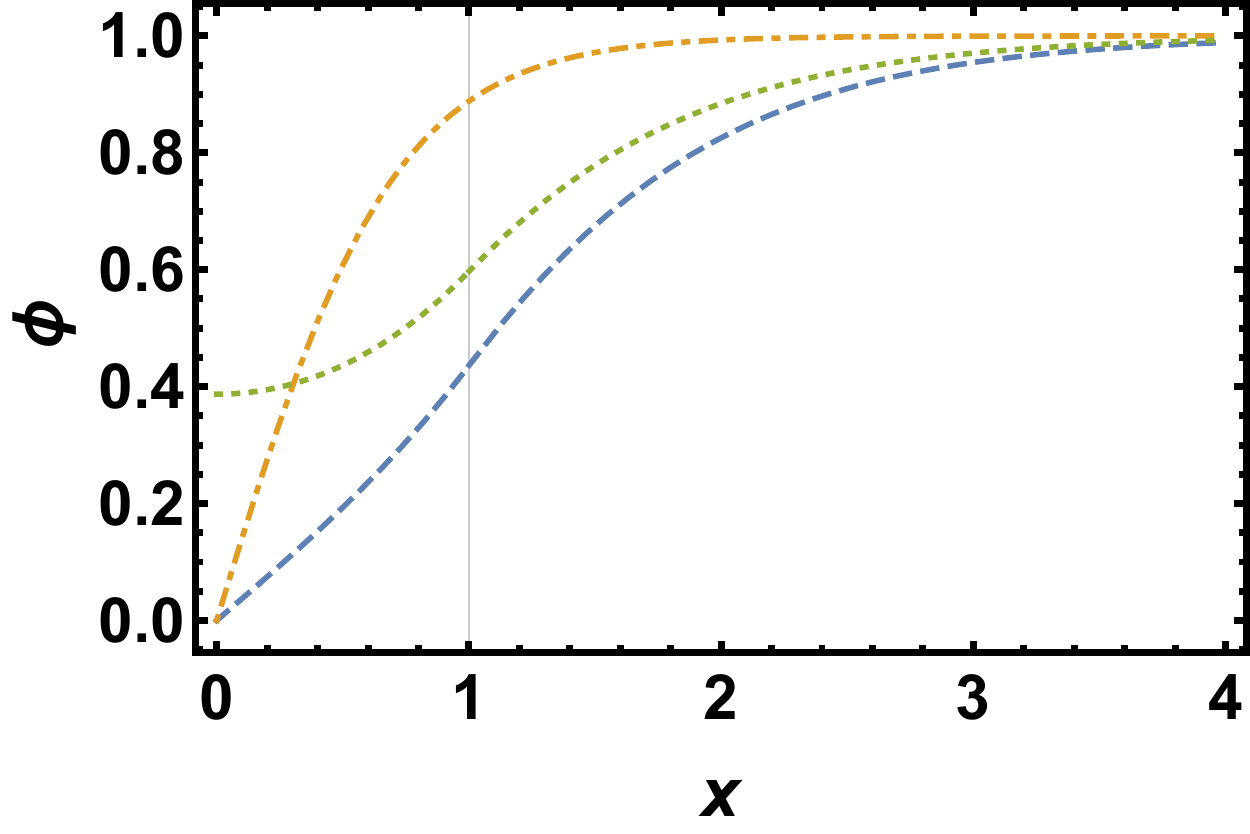}
\end{center}
\caption{Plot of the field profile for a symmetric slab (green dotted curve), 
an antisymmetric slab (blue dashed curve) and the vacuum domain wall (orange dot-dashed curve).
For the symmetric and antisymmetric slabs we take 
$L = M/\sqrt{\rho} = 1/\mu = 1$ and $\phi_0 = 1$. For the vacuum domain wall we take $1/\mu = 1$ and
$\phi_0 = 1$. Note that 
$\phi = 0$ at $x = 0$ for both the vacuum and matter
filled domain wall; $\phi$ rises more gradually for the matter filled domain wall.}
\label{fig:walls}
\end{figure}

To gain some insight into these solutions we note first that in the limit $L \rightarrow 0$ the symmetric
solution (\ref{eq:symmetricexact}) reduces to the vacuum $\phi = \phi_0$; the antisymmetric solution
(\ref{eq:antisymmetricexact}) reduces to the well known kink soliton \cite{tanmay}
\begin{equation}
\phi = \phi_0 \tanh \left[ \frac{\mu}{\sqrt{2}} x \right].
\label{eq:kink}
\end{equation}
Fig \ref{fig:walls} shows how the vacuum and the domain wall are modified by the presence of the slab.
We see that a test particle placed near a slab will be attracted to it by both gravity and the fifth force.
However for a given density and slab thickness the fifth force is greater if the slab is a domain wall. 
Although domain walls can exist without any matter since they attract matter we expect that in a
cosmological or astrophysical context domain walls will accrete matter. 

\subsection{The Analogy}

\label{sec:analogy}

The equations for the symmetron model are nonlinear partial differential equations and are 
difficult to solve in general. Under appropriate conditions
the electrostatic analogies that we develop in this paper
provide a simple approximation that facilitates calculation of the
symmetron profile and the forces on dense bodies. The analogies allow us to draw upon our intuition 
about electrostatics and thereby provide qualitative insights into symmetron gravity. 
Applications of the analogy are presented in section \ref{sec:image} and 
Appendix \ref{sec:applications}.

\subsubsection{Thin shell approximation}
\label{sec:thin}

In the thin shell electrostatic approximation we take $\phi = 0$ inside the dense body. In other words
the dense body is treated as an equipotential like a conductor in electrostatics. Outside the dense
body the symmetron field is assumed to obey Laplace's equation,
$\nabla^2 \phi = 0$. In addition the exterior symmetron field is assumed to obey the
boundary conditions that $\phi = 0$ on the surface of the dense body and
$\phi \rightarrow \phi_0$ far from the dense body. The justification for taking $\phi \approx 0$
is that zero is the equilibrium value of the field inside a dense body and it is presumed that the
symmetron field relaxes to that value everywhere except in a thin shell of thickness
$M/\sqrt{\rho}$ near the surface. Thus we are assuming that the shell thickness, $M/\sqrt{\rho}$ is small 
compared to the dimensions of the dense body. In the exterior it is really more accurate to take
the symmetron to obey $\nabla^2 \phi + \mu^2 \phi = 0$. However if $1/\mu$ is large
compared to all experimental scales of interest then it is acceptable to work with 
Laplace's equation $\nabla^2 \phi = 0$.

Within the approximations described in the preceding paragraph
the symmetron force on a dense body is determined by the gradient
of the symmetron field on its surface. The force is given by
\begin{equation}
{\boldsymbol {\cal F}} = - \frac{1}{2} \oint da \; (\hat{{\mathbf n}} \cdot \nabla \phi )^2 \hat{{\mathbf n}}.
\label{eq:forceanalog}
\end{equation}
Here $da$ is an area element on the surface of the dense body and 
$\hat{{\mathbf n}}$ denotes the outward pointing unit vector normal to the surface element. 
Eq (\ref{eq:forceanalog}) may be justified as follows.
Consider a small area element on the surface of the dense body. In the thin shell limit the length scale $M/\sqrt{\rho}$ is
much smaller than the local radius of curvature. Hence we may treat the interface as essentially flat locally. It follows
from our analysis of the semi-infinte slab in section \ref{sec:semislab} 
that the symmetron field must decay exponentially with depth as we descend into the dense body and 
the area element will be subject to a pressure equal in magnitude to $(1/2) ( \hat{{\mathbf n}} \cdot \nabla \phi )^2$
in accordance with eq (\ref{eq:thinpressure}). 
The pressure is directed inward leading to eq (\ref{eq:forceanalog}). 
It follows
readily from eq (\ref{eq:forceanalog}) that the torque on a dense body due to the symmetron field
is given by
\begin{equation}
{\boldsymbol \tau} = - \frac{1}{2} \oint da \; (\hat{{\mathbf n}} \cdot \nabla \phi )^2 ( {\mathbf r} \times \hat{{\mathbf n}} ).
\label{eq:torqueanalog}
\end{equation}
Here ${\mathbf r}$ is the position of the area element $da$.

It is instructive to compare the thin shell regime for the chameleon and the symmetron models. 
For the former the potential of a dense body is determined by the density of the body as well as the density
of the ambient medium whereas for the symmetron the potential is zero. Furthermore for the chameleon
the thickness of the shell is determined by the field gradient on the surface of the dense body \cite{jsf}. In contrast
for the symmetron the thickness of the shell is determined by the density of the body as noted above.

\subsubsection{Thick shell approximation}

In the opposite thick shell limit the symmetron field remains close to the vacuum value, $\phi \approx \phi_0$,
both inside and outside the dense objects. To leading order we may therefore replace $\phi \rightarrow \phi_0$
on the right hand side of eq (\ref{eq:scalar}). We find then that in the interior of the dense body the symmetron
field obeys Poisson's equation 
\begin{equation}
\nabla^2 \phi = \frac{\phi_0}{M^2} \rho.
\label{eq:thickpoisson}
\end{equation}
In the exterior the symmetron field obeys Laplace's equation $\nabla^2 \phi = 0$ and the boundary condition 
$\phi \rightarrow \phi_0$ far from the dense body. Thus dense bodies with thick shells are
analogous to charged insulators whereas dense bodies with thin shells are analogous
to perfect conductors. The symmetron force on a test particle of mass $m_0$, which is given
by ${\mathbf F}_\phi = - (m_0/M^2) \phi \nabla \phi$, simplifies to ${\mathbf F}_\phi = - (m_0 \phi_0/M^2) \nabla \phi$
if the symmetron field remains close to the vacuum value $\phi_0$. Thus in the thick shell limit both the
symmetron field equations and the force law are of the same form as the corresponding equations of 
Newtonian gravity. Moreover if we also assume, following Hinterbichler and Khoury \cite{hk}, 
that $\phi_0 M_P / M^2 \sim 1$,
then the symmetron fifth force is not only of the same form but is also comparable in strength to gravity. 

\subsection{An image problem}

\label{sec:image}

A small dense body in the thick shell limit does not by itself significantly
disturb the symmetron field. However if the small dense body is placed close to a larger dense
body with a thin shell, the small body does have a significant effect, analogous the the formation of an image
charge in electrostatics. A surprising twist is that in this case the image mass is negative so the
fifth force between the source particle and the image is repulsive. 

As a concrete instance consider a dense sphere of radius $R$ with a thin shell placed at the origin.
A small dense body of mass $m$ in the thick shell regime is placed on the $z$-axis at a distance
$r$ from the center of the thin shell sphere. Treating the small dense body as a point mass we wish 
to solve
\begin{equation}
\nabla^2 \phi = \frac{ \phi_0 }{M^2} m \delta (r - z) \delta (x) \delta (y)
\label{eq:imagediffeq}
\end{equation}
outside the thin-shell sphere subject to the boundary conditions $\phi = 0$ for $r = R$ 
and $\phi \rightarrow \phi_0$ 
for $r \rightarrow \infty$. Drawing upon the method of images \cite{jackson}
we see that the field in the
exterior of the thin-shell sphere will be that of three point masses. The first is $m$ the physical
object that is actually present. The second, $m' = - m R/r$,
is located along the $z$-axis at a distance $r' = R^2/r$ from the origin; the third,
$m'' = 4 \pi R M^2$ is located at the origin. Thus the field in the exterior of the thin-shell
sphere is
\begin{eqnarray}
\phi & = & \phi_0 \left( 1 - \frac{R}{s} \right) 
\nonumber \\
& + &
\frac{4 \pi \phi_0 m}{M^2} \left[
\left( \frac{s r^2}{R^2} + R^2 - 2 z r \right)^{-\frac{1}{2}} -
\left( s^2 + r^2 - 2 r z \right)^{-\frac{1}{2}} \right].
\nonumber \\
\label{eq:imagefield}
\end{eqnarray}
Here $s$ is the distance from the origin of the point at which the field is being evaluated
and $z$ is the $z$-coordinate of that point. 
The first term in eq (\ref{eq:imagefield}) 
is the field of the thin-shell sphere left to itself (or equivalently the field of the image 
mass $m''$), 
the positive term in the 
second line is the field of the image mass $m'$ and the final negative term is
the field of the point mass by itself.

In principle one can calculate the force between the point mass and the thin shell
sphere using the field in eq (\ref{eq:imagefield}) and the force formula
(\ref{eq:forceanalog}) but in practice it is easier to calculate the force between
the point mass and its two images. In the thick shell regime the fifth force between
two point masses $m_1$ and $m_2$ separated by a distance $r$ is given by
$F_{\phi} = (\phi_0^2/M^4) m_1 m_2 / (4 \pi r^2)$. Hence the force between the
point mass and the thin shell sphere is
\begin{equation}
F_\phi = \frac{m \phi_0^2 }{M^2} \frac{R}{r^2} - \frac{m^2 \phi_0^2}{4 \pi M^4} \frac{Rr}{(r^2 - R^2)^2}.
\label{eq:imageforce}
\end{equation}
If we compute $F_\phi/m$ and take the limit $m \rightarrow 0$ we recover the result for a test
mass near a thin shell sphere given by Hinterbichler and Khoury \cite{hk} and given 
in section \ref{sec:exactsphere}. 
For large $r$ the first term in eq (\ref{eq:imageforce}) dominates and the force is 
attractive but as $r \rightarrow R$ eventually the second repulsive term will dominate making the
net fifth force repulsive. There is no counterpart to this remarkable effect in Newtonian gravity.

Since repulsion is an unexpected outcome we briefly delineate the conditions under
which it might be expected to arise. It is clear from eq (\ref{eq:imageforce}) 
that the repulsive term dominates when the point mass is brought
close to the sphere. It is plausible that the electrostatic analogy holds only so long as the
point mass is at a distance much greater than the shell thickness. 
Placing the point mass at a distance that is a multiple of the shell thickness above the
dense sphere, eq (\ref{eq:imageforce}) reveals that the force is repulsive if the point mass $m$ is greater than
\begin{equation}
m > \frac{M^4}{R \rho}.
\label{eq:lower}
\end{equation}
There is also an upper bound that the point mass must respect for the following reason.
Let $a$ denote the radius of the test mass. (i) We need $a \ll R$ in order that we may 
consider it point-like. (ii) We also need $a < r - R$ in order that we can place the
point mass above the dense sphere. Evidently if we take $a \sim M / \sqrt{\rho}$ we can
easily meet both requirements. 
(iii) Finally for the
point mass itself to be in the thick shell regime we need $m/ a M^2 \ll 1$. Combining these conditions 
yields
\begin{equation}
m < \frac{M^4}{R \rho} \left( R \frac{ \sqrt{\rho} }{M} \right).
\label{eq:upper}
\end{equation}
Comparing eqs (\ref{eq:lower}) and (\ref{eq:upper}) and recalling that
$R \gg M/\sqrt{\rho}$ we see that there is a window
of values of the mass $m$ for which repulsion is obtained. 
According to ref \cite{upadhye} table top fifth force measurements 
are sensitive to the symmetron model with $M \sim 1$ TeV, a scale that is 
relevant to physics beyond the standard model. Choosing this value of $M$ and
taking $R \sim 1$ m and $\rho \sim 1000$ kg/m$^3$, we find that the 
shell thickness is of order 0.1 mm and 
test mass should have a mass between 0.1 mg to 1 kg
in order to experience a repulsive symmetron force \cite{footnote}. 

Naively one might assume that a force mediated by a scalar has to be attractive. 
This is certainly true for a linear theory but in fact there is no reason that this has to be true for
an interaction mediated by a non-linear interacting scalar field. The authors of ref \cite{gubser} have 
carefully investigated this question and been able to prove a remarkable but highly restricted 
theorem that for a pair of identical objects the force must be attractive; but their analysis cannot be 
generalized to the case of two highly asymmetric objects such as 
the large dense body and the small test mass considered above. The analysis above
by the method of images is highly suggestive and provides an intuitive explanation of how repulsive forces might
arise in this case, but it is not a rigorous proof that there is a regime in which repulsion is obtained. 
Further evidence in the form of bounds on the corrections to the electrostatic approximation, numerical simulations
or exact analysis of a simpler geometry are necessary to conclusively demonstrate the 
existence of a repulsive fifth force.


\section{Conclusion}

\label{sec:conclusion}

The electrostatic analogies presented in this paper show that in the unscreened thick shell
regime the symmetron field behaves much like Newtonian gravity. However in the screened
thin shell regime relevant to the solar system and the laboratory it exhibits effects unlike
any shown by Newtonian gravity. These effects can be tested on scales ranging from the
astrophysical to table-top atom interferometry \cite{mueller,hinds,burrage}. Although the electrostatic analogy is a useful source for qualitative
insight, in order to make quantitative contact with observations and experiments, numerical work will be necessary. 
For example, in order to design atom interferometry experiments
that exploit the enhancement of the field around an ellipsoid, it is necessary to obtain
an accurate numerical solution that takes into account the walls of the vacuum chamber, analogous
to the computations that have been carried out for spheres within the chameleon model in ref \cite{elder}.

Much of what is known about chameleon and symmetron gravity in the dynamic and relativistic
regimes is derived from numerical simulations (see ref \cite{trodden} and references therein). 
The authors of ref \cite{dani} have studied spherical collapse
in the chameleon model, radial dynamics has been studied perturbatively in \cite{silvestri} and the issue of monopole radiation in modified gravity theories is taken up in \cite{monopole}.
Further work along these lines 
is desirable as better qualitative understanding may help guide numerical simulations 
and suggest new observational tests.

\appendix
\section{Applications of the Electrostatic Analogy}

\label{sec:applications}

\vspace{3mm}


\subsection{Uniform Dense Sphere}
As a check let us re-calculate the symmetron field profile for a spherical body of radius $R$ and density $\rho$
using the thin-shell electrostatic approximation outlined above. We take $\phi = 0$ for $r < R$. In the exterior of
the dense body we assume that $\phi$ depends only on the radial co-ordinate $r$ and hence Laplace's 
equation has the form
\begin{equation}
\frac{1}{r} \frac{d^2}{d r^2} ( r \phi ) = 0 \hspace{3mm} {\rm for} \hspace{3mm} r > R.
\label{eq:sphericallaplacian}
\end{equation}
Eq (\ref{eq:sphericallaplacian}) has the most general solution
\begin{equation}
\phi = A + \frac{B}{r}
\label{eq:gensol}
\end{equation}
where $A$ and $B$ are arbitrary coefficients. Imposing the boundary conditions that
$\phi \rightarrow \phi_0$ as $r \rightarrow \infty$ and $\phi = 0$ for $r = R$
we recover eq (\ref{eq:thinshellsphereelectrostatic}) derived previously as a limiting
case of the result given by Hinterbichler and Khoury \cite{hk}. 

Similarly in the thick shell approximation we obtain the interior solution 
\begin{equation}
\phi = A + \frac{1}{6} \frac{ \rho \phi_0 }{M^2} r^2 
\label{eq:thickinterior}
\end{equation}
by solving eq (\ref{eq:thickpoisson}).
Here we have made use of the condition that $\phi$ must be regular at the origin. 
We obtain the exterior solution
\begin{equation}
\phi = \phi_0 + \frac{B}{r}
\label{eq:thickexterior}
\end{equation}
by solving Laplace's equation subject to spherical symmetry and the boundary condition
$\phi \rightarrow \phi_0$ as $r \rightarrow \infty$. The coefficients $A$ and $B$ can be 
determined by by matching $\phi$ and $d \phi/ dr$ at $ r = R$ yielding 
eqs (\ref{eq:thickshellsphere}) and (\ref{eq:thicksphereelectrostatic}) derived previously by taking 
limiting cases of the result derived by Hinterbichler and Khoury \cite{hk}. 

\subsection{Dense Sphere in a Uniform Gradient}

We turn now to problems that would be difficult to analyze without the electrostatic analogy.
In this section we analyze the force on a dense sphere placed in a symmetron field with a 
uniform gradient, $\nabla \phi = F \hat{{\mathbf z}}$. The force on a test particle placed
in this field would be ${\boldsymbol {\cal F}} = (m_0/M^2) \phi \nabla \phi$. If we naively
assume that the dense body has no effect on the symmetron field we may treat it as
a test particle. Setting $m_0 \rightarrow 4 \pi \rho R^3/3$, $\phi \rightarrow \phi_0$ and
$\nabla \phi = F \hat{{\mathbf z}}$ we obtain
\begin{equation}
{\boldsymbol {\cal F}} = \frac{4 \pi}{3} \rho R^3 \frac{\phi_0 F}{M^2} \hat{{\mathbf z}}.
\label{eq:thicksphereforce}
\end{equation}
This formula should describe the force on a dense sphere with a thick shell. However, as we
will now show, it greatly overestimates the force on a dense sphere with a thin shell. In the thin
shell limit the symmetron field is greatly modified by the presence of the dense body exactly as
a conducting sphere placed in a uniform electric field modifies the electric field in its vicinity. 
As a result the force on the thin shell sphere is suppressed by a factor $1/\alpha$, exactly
the same factor by which the force that the sphere would exert on a test particle is suppressed. 

In order to calculate the force we must first calculate the field profile near the dense sphere.
In the thin-shell electrostatic approximation we set $\phi = 0$ in the interior of the sphere
(for $r < R$). In the exterior the field obeys Laplace's equation and the
modified boundary condition that $\phi \rightarrow \phi_0 + F r \cos \theta$ far from the dense
body. This problem is mathematically identical to the textbook electrostatic problem of a
conducting sphere placed in a uniform electric field and we can draw upon that analysis here.
We take the exterior field to have the form
\begin{equation}
\phi = A + \frac{B}{r} + C r \cos \theta + \frac{D}{r^2} \cos \theta.
\label{eq:dipoleexpansion}
\end{equation}
where $A, B, C$ and $D$ are arbitrary coefficients determined by the boundary
conditions. Imposing those conditions we obtain
\begin{equation}
\phi = \phi_0 \left( 1 - \frac{R}{r} \right) + F r \cos \theta - F \frac{R^3}{r^2} \cos \theta
\label{eq:uniformexterior}
\end{equation}
for $r > R$ and $\phi = 0$ for $r < R$. 

It is now straightforward to calculate the force on the dense
body making use of eqs (\ref{eq:forceanalog}) and (\ref{eq:uniformexterior}). Evaluating
the integral over the surface of the sphere yields
\begin{equation}
{\boldsymbol {\cal F}} = 4 \pi \phi_0 F R \hat{{\mathbf z}}.
\label{eq:thinsphereforce}
\end{equation}
This is smaller than the naive estimate (\ref{eq:thicksphereforce}) by a factor of $1/3 \alpha$.
Physically the reason is that the dense body acts like a conductor in the thin shell
limit. The symmetron field
is essentially constant in its interior except in a thin shell near the surface where there is a small
gradient that leads to the residual force calculated here. In the thick shell limit the symmetron field
would have an essentially uniform gradient throughout the interior of the dense body leading
to a much larger force. Remarkably the force on the thin shell sphere (\ref{eq:thinsphereforce})
is independent of the mass of the body and is determined entirely by its volume. 

For comparison we present the corresponding results for the chameleon which 
were omitted in ref \cite{jsf} for the sake of brevity. A uniform sphere of radius $R$ and density
$\rho_c$ placed in an ambient medium of density $\rho_\infty$ with a chameleon gradient
$\nabla \phi = F \hat{{\mathbf z}}$ experiences a force given by eq (\ref{eq:thinsphereforce}) but
with the replacement $\phi_0 \rightarrow \phi_\infty - \phi_c$. Here $\phi_\infty$ and $\phi_c$
are the equilibrium values of the chameleon field in a homogeneous medium of density
$\rho_\infty$ and $\rho_c$ respectively. Thus the chameleon force on a sphere in the thin
shell regime depends not only on the volume but also the mass of the sphere as well as the
ambient density. It is reduced compared to the force in the thick shell regime by a factor
of $1/3 \alpha'$ where  $\alpha' = \delta/R = M_{{\rm P}} (\phi_\infty - \phi_c)/
\beta \rho_c R^2$ is the chameleon
thin shell factor.


\subsection{Dense Ellipsoid}
\label{sec:ellipsoid}

It is well known in electrostatics that the electric field near the surface
of a conductor is greatly enhanced near sharp corners. This is sometimes called the lightning rod
effect. By virtue of the electrostatic analogy it follows that the symmetron field will show a
similar enhancement. The Newtonian gravitational field by contrast shows no such enhancement. 
In order to illustrate the lightning rod effect for symmetron forces we analyze the field near an
ellipsoidal dense body. We find that the resulting symmetron gradient is highly anisotropic and 
enhanced near the pointed ends of the ellipsoid. There are close parallels to the chameleon and
we follow the corresponding analysis of Jones-Smith and Ferrer \cite{jsf}. 

 
In the following it will be convenient to use prolate spheroidal
co-ordinates $(\xi, \eta, \varphi)$ that are related to Cartesian co-ordinates $(x,y,z)$ via
$\xi = (r_+ + r_-)/a$; $\eta = (r_+ - r_-)/a$; and $\varphi = \tan^{-1} (y/x)$.
Here $r_{\pm} = \sqrt{ x^2 + y^2 + (z \pm a/2)^2 }$ denotes the distance to the two foci 
located on the $z$-axis symmetrically about the origin and a distance $a$ apart. The co-ordinates
lie in the range $\xi \geq 1$, $-1 \leq \eta \leq 1$ and $0 \leq \varphi < 2 \pi$. Surfaces of constant
$\xi$ are ellipsoids with major axis $a \xi$ and eccentricity $1/\xi$.
Points on a surface of fixed $\xi$ are distinguished by the values of $\eta$ and $\varphi$ which
are essentially the ``latitude'' and ``longitude'' on the surface of the ellipsoid. $\eta = \pm1$
corresponds to the poles and $\eta = 0$ corresponds to the equator. Some further useful information
about this co-ordinate system is collected in Appendix \ref{sec:prolate}. 

The surface of the dense body is assumed to be an ellipsoid defined
by $\xi = \xi_0$. 
Hence the major axis of the ellipsoid is $\frac{1}{2} a \xi_0$
and the minor axis is $\frac{1}{2} a \sqrt{ \xi_0^2 - 1}$. It is convenient to define an equivalent
radius $R_c$ so that the volume of the ellipsoid is given by $\frac{4}{3} \pi R_c^3$. Thus
$R_c = \frac{1}{2} a [ \xi_0 (\xi_0^2 - 1) ]^{1/3}$. 
Within the thin-shell electrostatic analogy 
we wish to solve $\nabla^2 \phi = 0$ in the exterior of the dense body subject to the boundary
conditions $\phi = 0$ at $\xi = \xi_0$ and $\phi \rightarrow \phi_0$
as $\xi \rightarrow \infty$. The solution to this problem is given by Morse and Feshbach \cite{mf} as
\begin{equation}
\phi = \phi_0 - \phi_0 \frac{ Q_0^0(\xi) }{Q_0^0 (\xi_0)}.
\label{eq:mfsol1}
\end{equation}
Here $Q_0^0 (\xi ) = \frac{1}{2} \ln [ ( \xi  + 1 )/(\xi - 1) ]$. 

Further insight into this result is obtained by examining the symmetron potential 
far from the ellipsoid ($\xi \gg \xi_0$). 
In this regime, if we adopt spherical polar co-ordinates, the potential has the form of a multipole expansion,
\begin{equation}
\phi \approx \phi_0 - \left[ \frac{ a \phi_0 }{2 Q_0^0(\xi_0)} \right] \frac{1}{r} -
\left[ \sqrt{ \frac{ 4 \pi }{5 }} \frac{a^3 \phi_0}{24 Q_0^0(\xi_0) } \right] Y_{20} (\theta, \varphi) \frac{1}{r^3} + \ldots
\label{eq:ellipsoidalmultipole}
\end{equation}
If we keep only the leading isotropic term in the potential 
we obtain
\begin{equation}
\phi \approx \phi_0 - \phi_0 \frac{R_c}{r} f(\xi_0) + \ldots
\label{eq:shape}
\end{equation}
Here we have eliminated $a$ in favor of $R_c$ so that the $\xi_0$ dependence provides the
shape dependence of the field at a fixed ellipsoid volume. Thus 
\begin{equation}
f(\xi_0) = \frac{1}{\xi_0^{1/3} (\xi_0^2 - 1)^{1/3} Q_0^0 (\xi_0) }
\label{eq:shapefactor}
\end{equation}
is a factor that depends on the shape of the ellipsoid. Recall that $1/\xi_0$ is the eccentricity of the ellipsoid and
hence $\xi_0 \rightarrow \infty$ corresponds to a sphere and $\xi_0 \rightarrow 1$ to an extremely
elongated ellipsoid that essentially forms a sheath for the line segment joining the foci. 
In these limits $f \rightarrow 1$ and $f \rightarrow \infty$ respectively.
The divergence is spurious as discussed below 
but it is clear that for a fixed volume (or equivalently
fixed $R_c$) the far field of an elongated ellipsoid is stronger than that of one that is essentially spherical. 
By contrast the isotropic term in the gravitational far field is 
\begin{equation}
\psi = - \frac{1}{6} \frac{ \rho R_c^3}{M_{{\rm P}}^2} \frac{1}{r} + \ldots
\label{eq:isotropicgrav}
\end{equation}
It has no dependence on the shape at all. 

For the thin shell electrostatic approximation to apply we need $R_c$ to be much
greater than the shell thickness $\sqrt{\rho}/M$. For a sufficiently elongated ellipsoid
we also need to impose that the radius of curvature at the poles should exceed the
shell thickness. The radius of curvature at the poles is given by $R_c (\xi_0^2 - 1)^{2/3}/\xi_0^{4/3}$.
Hence we need 
\begin{equation}
\xi_0 \geq \frac{ 1 }{ \sqrt{ 1 - (1/\alpha)^{3/4} } } \approx 1 + \frac{1}{2} \left( \frac{1}{\alpha} \right)^{3/4}.
\label{eq:maxxi}
\end{equation}
Here $\alpha = \rho R_c^2/M^2$ and the approximate equality holds when $\alpha \gg 1$. 
Thus for large $\alpha$ the electrostatic approximation holds quite close to $\xi_0 \approx 1$.
It follows from eq (\ref{eq:maxxi}) that the maximum value attained by the shape factor before
the electrostatic approximation breaks down is 
\begin{equation}
f_{{\rm max}} \approx \frac{ 4 (\alpha)^{1/4} }{ \ln (2 \alpha) }.
\label{eq:maxfactor}
\end{equation}
Recall that the thin shell reduction factor for a sphere is $1/\alpha$. Hence although the enhancement factor
(\ref{eq:maxfactor}) can be substantial it cannot completely offset the thin shell reduction. 

The lightning rod effect is manifested in the near field also as an enhancement
of the field due to an ellipsoid 
compared to that of a sphere of the same volume. The enhancement is stronger at the 
the poles than it is near the equator. To demonstrate this effect we compute the gradient
of the symmetron potential on the surface of the ellipsoid $\xi = \xi_0$ making use of eqs (\ref{eq:mfsol1}), 
(\ref{eq:gradient}) and (\ref{eq:scalefactors}). We find
\begin{equation}
\nabla \phi = \frac{\phi_0}{R_c} \frac{ \xi_0^{1/3} }{ (\xi_0^2 - 1)^{1/6} ( \xi_0^2 - \eta^2 )^{1/2} Q_0^0 (\xi_0) }
\hat{{\mathbf n}}_\xi.
\label{eq:neargrad}
\end{equation}
Here $\hat{{\mathbf n}}_\xi$ is the outward pointing unit vector perpendicular to ellipsoidal surfaces
of constant $\xi$. We have eliminated $a$ in favor of $R_c$ using eq (\ref{eq:rc}) so that the $\xi_0$ dependence
provides the shape dependence of the surface field gradient at fixed ellipsoidal volume. 


As a first application one can check that in the limit $\xi_0 \rightarrow \infty$
eq (\ref{eq:neargrad}) simplifies to $\nabla \phi = (\phi_0/R_c) \hat{{\mathbf r}}$, the
expected result for the chameleon gradient on the surface of a thin shell sphere 
(\ref{eq:thinshellsphereelectrostatic}). It is more interesting to compute $\nabla \phi$ 
at the equator and the poles of an ellipsoid with maximal elongation given by
eq (\ref{eq:maxxi}). We find that at both the poles and the equator the gradient is given by
\begin{equation}
\nabla \phi = \frac{8}{3} \frac{\phi_0}{R_c} \frac{ \alpha^\nu}{ \ln \alpha + \frac{4}{3} \ln 4 } \hat{{\mathbf n}}_\xi
\label{eq:nearenhancement}
\end{equation}
where the exponent $\nu = 1/8$ for the equator and $\nu = 1/2$ at the poles. 
Recall the thin shell reduction factor for a dense sphere is $1/\alpha$. 
Hence although there is an enhancement of the field at both the equator and the poles
compared to a sphere of the same volume, elongation does not completely offset the
thin shell reduction even at the poles where the enhancement is greater than at the equator.


The enhancement in the near field computed here is relevant to experiments
that measure the local acceleration of atoms using interferometry \cite{mueller, hinds, burrage}. However in
order to make quantitative contact with such experiments it is not sufficient
to analyze an isolated ellipsoid due to the proximity in experiments of the vacuum chamber walls.
This analysis is left open for future work. 

\subsection{Torque on a Dense Ellipsoid in a uniform gradient}
\label{sec:torque}

We now consider a dense ellipsoid with a thin shell placed in a symmetron field with a uniform gradient.
Remarkably the ellipsoid will experience a torque that tends to align its major axis with the applied
symmetron field gradient. This is analogous to the behavior of a conducting ellipsoid placed in a uniform
electric field. No such effect occurs for a dense body placed in a uniform gravitational field or for an ellipsoid
with a thick shell. The physical reason for the torque is most easily understood in terms of the conductor analogy. 
The applied electric field causes the charges on the conductor to rearrange and form a dipole. Since the charges
preferentially localize near the poles of the ellipsoid the dipole moment is aligned with the major axis of the 
ellipsoid rather than the applied electric field. The torque then arises because an electric dipole placed in an
electric field experiences a torque that tends to align the dipole with the electric field. The analogous 
effect for the chameleon was pointed out by Jones-Smith and Ferrer \cite{jsf} 
and we closely follow that analysis.

We assume that the ellipsoid is aligned with the $z$-axis and the applied symmetron gradient in which it is
placed lies in the $x$-$z$ plane and makes an angle $\gamma$ to the $z$-axis. Thus we wish to solve
Laplace's equation subject to the boundary conditions that $\phi = 0$ on the surface of the ellipsoid ($\xi = \xi_0$)
and that $\phi \rightarrow \phi_0 + F z \cos \gamma + F x \sin \gamma$ far from the ellipsoid ($\xi \rightarrow 
\infty$). The solution has the form
\begin{equation}
\phi = \phi_1 + \phi_2.
\label{eq:superpos}
\end{equation}
Here $\phi_1$ is given by the right hand side of eq (\ref{eq:mfsol1}) and $\phi_2$ by
\begin{eqnarray}
\phi_2 & = & - \frac{F R_c}{ [ \xi_0 ( \xi_0^2 - 1) ]^{1/3} }
\{ (\cos \gamma) \eta \left[ \frac{ \xi_0 }{ Q_1^0 (\xi_0) } Q_1^0 (\xi) - \xi \right] 
 \nonumber \\
& + & \sin \gamma \cos \varphi \sqrt{1 - \eta^2} \left[ \frac{ \sqrt{ \xi_0^2 - 1 } }{Q_1^1 (\xi_0) } Q_1^1 (\xi) - 
\sqrt{ \xi^2 - 1 } \right] \} \nonumber \\
\label{eq:mfsol2}
\end{eqnarray}
with $Q_0^1 = \xi Q_0^0 (\xi ) - 1$ and 
\begin{equation}
Q_1^1 (\xi) = \sqrt{ \xi^2 - 1 } \left[ \frac{ \xi }{ \xi^2 - 1 } - Q_0^0 (\xi ) \right].
\label{eq:q11}
\end{equation}
This solution is constructed by adapting the solution given by Morse and Feshbach \cite{mf} of the
potential around a grounded conductor placed in a uniform electric field. 

The torque may then be readily calculated by use of eq (\ref{eq:torqueanalog}). The only non-vanishing 
component is 
\begin{equation}
\tau_y = \pi R_c^3 F^2 (\sin \gamma \cos \gamma) \alpha (\xi_0)
\label{eq:torque}
\end{equation}
where $\alpha$ is a shape dependent factor given by
\begin{equation}
\alpha (\xi_0) = \frac{ \frac{2}{3} + 2 ( 1 - \xi_0 ) Q_1^0 (\xi_0) }{Q_1^0 (\xi_0) Q_1^1 (\xi_0) 
\xi_0^{5/3} ( \xi_0^2 - 1 )^{3/2} }.
\label{eq:torqueshape}
\end{equation}
Qualitative features of this result can be understood by use of the electrostatic analogy. 
For a conductor one would expect the torque to be proportional to the square of the electric field
since the dipole moment itself is induced by the electric field and is proportional to it. This accounts
for the $F^2$ dependence in eq (\ref{eq:torque}). The dependence on $\gamma$ reflects the fact that
the dipole moment is determined by the electric field along the major axis; the torque by the component
perpendicular to the major axis. The magnitude of the dipole moment is proportional to $R_c^3$ because
the charge on either hemisphere is proportional to the area which scales as $R_c^2$ and the distance
between the charges brings in an additional factor of $R_c$. Finally we expect that for a sphere there
should be no torque and indeed the shape factor $\alpha \rightarrow 0$ as $\xi_0 \rightarrow \infty$ (to be precise,
$\alpha \approx 3/\xi_0^{2/3}$ for $\xi_0 \gg 1$). 

In the thin shell regime generally fifth force effects are overwhelmed by gravity. 
It is therefore remarkable that if an ellipsoid is placed in a uniform gravitational and symmetron field
only the symmetron field exerts a torque. Potentially this effect could serve as the basis of a torsion
balance experiment to constrain symmetron gravity.

\section{Prolate Spheroidal Co-ordinates}
\label{sec:prolate}

Prolate spheroidal co-ordinates were introduced in section \ref{sec:ellipsoid}. Here we 
collect some additional formulae that are useful for the calculations described in 
sections \ref{sec:ellipsoid} and \ref{sec:torque}. 
The distance between two nearby points is given by 
\begin{equation}
d s^2 = h_\xi^2 d \xi^2 + h_\eta^2 d \eta^2 + h_\varphi^2 d \varphi^2.
\label{eq:lineelement}
\end{equation}
The scale factors are 
\begin{eqnarray}
 h_\xi & = & \frac{a}{2} \sqrt{ \frac{ \xi^2 - \eta^2 }{ \xi^2 - 1 } }; \nonumber \\
h_\eta & = & \frac{a}{2} \sqrt{ \frac{ \xi^2 - \eta^2 }{ 1 - \eta^2 } }; \nonumber \\
h_\varphi & = & \frac{a}{2} \sqrt{ ( \xi^2 - 1 ) (1 - \eta^2) }. 
\label{eq:scalefactors}
\end{eqnarray}
The gradient of a scalar $f$ is 
\begin{equation}
\nabla f = \hat{{\mathbf n}}_\xi \frac{1}{h_\xi} \frac{\partial f}{\partial \xi} +
\hat{{\mathbf n}}_\eta \frac{1}{h_\eta} \frac{\partial f}{\partial \eta} +
\hat{{\mathbf n}}_\varphi \frac{1}{h_\varphi} \frac{\partial f}{\partial \varphi}.
\label{eq:gradient}
\end{equation}
The unit vector $\hat{{\mathbf n}}_\xi$ is given by
\begin{equation}
\hat{{\mathbf n}}_\xi = \eta \sqrt{ \frac{ \xi^2 - 1 }{ \xi^2 - \eta^2 } } \hat{{\mathbf z}} +
\xi \sqrt{ \frac{1 - \eta^2}{\xi^2 - \eta^2} } ( \cos \varphi \hat{{\mathbf x}} + \sin \varphi \hat{{\mathbf y}} ).
\label{eq:unitnormal}
\end{equation}
This vector is perpendicular to the ellipsoidal surfaces of fixed $\xi$ and points outward.
The area element on the surface of the ellipsoid $\xi = \xi_0$ is
\begin{equation}
da = h_\eta h_\varphi d \eta d \varphi.
\label{eq:areaelement}
\end{equation}
For calculation of the torque it is convenient to have explicit expressions for the cartesian
co-ordinates in terms of prolate spheroidal co-ordinates.
\begin{eqnarray}
z & =  & \frac{a}{2} \xi \eta, 
\nonumber \\
x & = & \frac{a}{2} \sqrt{ (\xi^2 - 1)(1 - \eta^2) } \cos \varphi,
\nonumber \\
y & = & \frac{a}{2} \sqrt{ (\xi^2 - 1)(1 - \eta^2) } \sin \varphi.
\label{eq:cartesian}
\end{eqnarray}
The scale of the ellipsoid $\xi = \xi_0$ is set by $a$ the interfocal distance. Sometimes it is convenient
to work with $R_c$ given by
\begin{equation}
R_c = \frac{a}{2} [ \xi ( \xi_0^2 - 1 ) ]^{1/3}.
\label{eq:rc}
\end{equation}
$R_c$ is defined as the radius of a sphere that has the same volume as the ellipsoid. 



\begin{thebibliography}{999}
\bibitem{footnote} As a practical matter
there is an additional constraint on $m$ and $a$ due to the fact that under laboratory
conditions the density of the test mass would at most be $2 \times 10^4$ kg/m$^3$. 
By a judicious choice of distance $r-R$ it is possible to meet this constraint and still
get repulsive forces for the representative values of $M, R$ and $\rho$ considered here.
\end{thebibliography}

\end{document}